\documentclass[12pt]{article}
\usepackage{setspace}
\usepackage{graphicx}
\usepackage{dcolumn}
\usepackage{bm}
\usepackage{times}
\usepackage{fullpage}
\usepackage{amsmath}
\begin{document}

\title{Enhanced and directional single photon emission in hyperbolic metamaterials}

\author{Ward D. Newman, Cristian L. Cortes and Zubin Jacob* \\
\small Department of Electrical and Computer Engineering, \\
\small University of Alberta, Edmonton, AB T6G 2V4, Canada \\
\small \emph{*zjacob@ualberta.ca}\\ \\
}
\date{}
\newcommand{\e}{\textrm{e}} 
\newcommand{\diff}[2] {\frac{\textrm{d}#1}{\textrm{d}#2}} 

\maketitle
\doublespacing 

\begin{abstract}
\noindent We propose an approach to enhance and direct the spontaneous emission from isolated emitters embedded inside hyperbolic metamaterials into single photon beams. The approach rests on collective plasmonic Bloch modes of hyperbolic metamaterials which propagate in highly directional beams called quantum resonance cones.  We propose a pumping scheme using the transparency window of the hyperbolic metamaterial that occurs near the topological transition. Finally, we address the challenge of outcoupling these broadband resonance cones into vacuum using a dielectric bullseye grating.  We give a detailed analysis of quenching and design the metamaterial to have a huge Purcell factor in a broad bandwidth inspite of the losses in the metal.  Our work should help motivate experiments in the development of single photon sources for broadband emitters such as nitrogen vacancy centers in diamond.  
\end{abstract}

\section{Introduction}

Nanoscale light-matter interactions can be tailored using plasmonic approaches to have an impact on quantum information processing \cite{nielsen2010quantum}. It has been conclusively demonstrated that propagating surface plasmon polaritons (SPPs) can possess non-classical properties such as entanglement \cite{altewischer2002plasmonassisted} and squeezing \cite{huck2009demonstration}. Furthermore, inspite of the decoherence expected from losses and electron collisions, these properties are preserved during propagation and are manifested in the outcoupled  photons \cite{dimartino2012quantum}. Thus SPPs could potentially be used as a carrier of quantum information at the nanoscale \cite{jacob2011plasmonics,jacob2012quantum}.

One important application where plasmonics can play a key role is in the development of room temperature single photon sources \cite{lounis2005singlephoton,grangier2004focus}. Isolated emitters like single dye molecules or quantum dots emit single photons however their efficiency is low for practical applications. The figures of merit for single photon sources are the collection efficiency which can be increased by directional spontaneous emission and the Purcell effect \cite{esteban2010optical,friedler2009solidstate,lee2011planar}. Deterministic out-coupling of single photons and tolerance to emitter positioning is another important factor affecting device performance \cite{englund2010deterministic}. 

At low temperatures when the linewidth of the emitter is narrow, resonant routes using microcavities can provide the Purcell enhancement. The broadband nature of many quantum emitters such as nitrogen vacancy (NV) centers in diamond make them unsuitable for conventional resonant cavity based approaches \cite{aharonovich2011diamond}. The high index of diamond also leads to total internal reflection of photons and a poor collection efficiency (4 \%) \cite{babinec2010diamond}. Therefore, the use of nanoplasmonic structures capable of a broadband Purcell effect is necessary for efficient extraction of single photons \cite{choy2011enhanced,huck2011controlled}. Plasmonic metamaterials which engineer the spontaneous emission can also be used to enhance the absorption spectrum of the isolated emitter \cite{tanaka2010multifold}. However, a significant limitation of any plasmonic approach is the presence of loss and non-radiative quenching \cite{ford1984electromagnetic}. Especially near the plasmon resonance, only a modest increase in quantum efficiency and collection efficiency is possible \cite{sun2007practicable}. The key is to move away from resonant approaches and focus on low mode volume plasmons. We emphasize that for applications requiring indistinguishable photons, photonic crystal and narrowband cavity approaches are ideally suited whereas plasmonic/metamaterial approaches have an advantage for broadband applications.

In this paper, we show that tailored plasmonic metamaterials support collective modes which can efficiently channel single photons from an isolated quantum emitter to highly directional beams and simultaneously provide a broadband Purcell effect. Our design for the single photon source, compatible with emitters such as NV centers in diamond, takes into account all non-idealities in the structure arising from the finite patterning scale, absorption and dispersion. We also provide a detailed account of quenching and function away from resonance to achieve an efficient single photon source inspite of the losses.

\section{Quantum Resonance Cones}
\begin{figure}
\begin{center}
\includegraphics[width=0.74\textwidth]{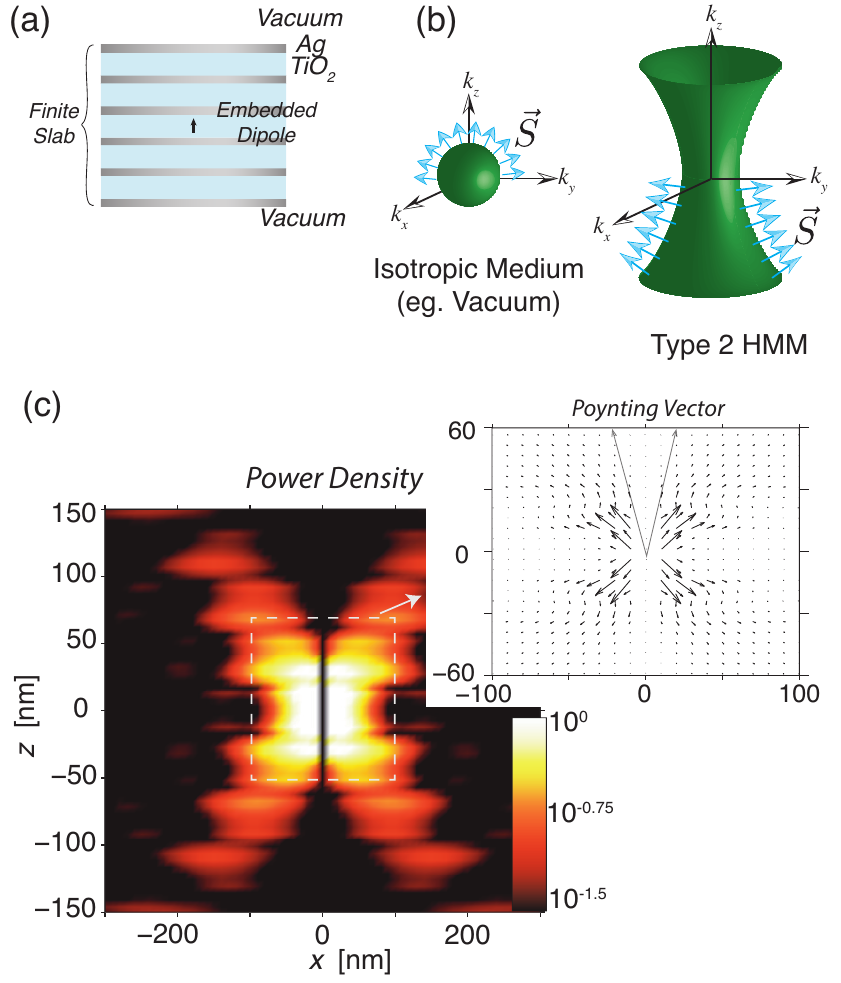}
\caption{(a) A multilayer stack of Ag/TiO$_2$ with 10/30 nm layer thicknesses behaves like a homogeneous metamaterial slab with hyperbolic dispersion ($\epsilon_\parallel<0,\epsilon_\perp>0$) above $\lambda \approx 720$ nm. The emitter is embedded symmetrically in a 30 nm layer of TiO$_2$. (b)  The direction of power flow is normal to the isofrequency surface. In HMMs, the power flow of all high-$k$ states tends to {\it bunch} and point in the same direction, thus forming resonance cones. (c) A  quantum dot embedded in a HMM emits preferentially into high-$k$ states within the resonance cone ($\lambda = 800$ nm). The Power Density and Poynting Vector are are normalized by the total time averaged power emitted from the oscillating point dipole. The grey arrows show the asymptotic direction of power flow for extremely high-$k$ states for an ideal HMM.}
\label{ResonanceConesFigure}
\end{center}
\end{figure}

Our approach relies on engineering the plasmonic Bloch modes of periodic metal-dielectric structures. Such 1D (multilayer) or 2D (nanowire) plasmonic crystals can behave as an effective metamaterial when the lattice spacing is far below the operating wavelength \cite{xiong2007twodimensional,kabashin2009plasmonic,elser2007nonlocal}. We consider a 1D multilayer metal-dielectric lattice as shown in Fig.\ref{ResonanceConesFigure}a. The short range propagating surface plasmon polaritons on each metal-dielectric interface couple leading to Bloch modes with unique properties not available in conventional photonic crystals. They are described by effective medium theory (EMT) in the metamaterial limit by a homogeneous medium that has a metallic dispersion along one direction but a dielectric dispersion in the perpendicular direction.

The multilayer structure behaves as an effective metamaterial slab with an extremely anisotropic dielectric tensor $\bar{\bar{\epsilon}}=\textrm{diag}[\epsilon_\parallel,\epsilon_\parallel,\epsilon_\perp]$,  where the directions are parallel and perpendicular to the above mentioned layers.  Extraordinary plane wave propagation in uniaxial anisotropic media is governed by the dispersion relation
\begin{equation}
k_x^2/\epsilon_\perp+k_z^2/\epsilon_\parallel = \left(\omega/c\right)^2
\label{dispersion}.
\end{equation}
which describes an open hyperboloid (Fig.\ref{ResonanceConesFigure}(b)) when $ \epsilon_\parallel \epsilon_\perp < 0$. These artificial media are known as hyperbolic metamaterials (HMM) \cite{smith2004partial}. They have the property of large momentum bulk propagating waves (high-$k$ modes with unbounded magnitudes of $k_x$ and $k_z$) which arise due to  surface-plasmon-polariton Bloch waves in the plasmonic crystal.  For ordinary materials such as glass  with $\epsilon_\parallel\epsilon_\perp>0$, the dispersion relation describes a bounded sphere and the magnitudes of $k_x$ and $k_z$ have an upper cut-off. Above this cut-off, waves are evanescent and simply decay away.

The Poynting vector in a medium is related to the normal vector to the dispersion relation (blue arrows in Fig.\ref{ResonanceConesFigure}(b)). The Poynting vector for various plane waves in the HMM, given by $\overrightarrow{S} = k_x/(k_0\epsilon_{\perp}) \hat{x}+k_z/(\epsilon_{\parallel}k_0) \hat{z}$, where $k_x$ and $k_z$ are related by the dispersion relation, eq. \ref{dispersion}, lie within a narrow region known as the resonance cone. Thus energy flow due to the bulk plasmonic Bloch modes in this metamaterial is inherently directional.  The half angle of the plasmonic resonance cone is given by  \cite{fisher1969resonance}
\begin{equation}
	\tan \theta_{RC} = \sqrt{-\frac{\epsilon_\parallel}{\epsilon_\perp}}.
\end{equation}
Furthermore, for all the waves in the medium with wavevectors along the asymptotes of the hyperbola, the Poynting vectors point in the same direction. Since there are infinitely many waves (in the EMT limit) with wavevectors along the asymptotes of the hyperbola, a spatial crowding of Poynting vectors is expected in preferred directions (Fig.\ref{ResonanceConesFigure}(b)).
This phenomenon has been observed in anisotropic plasmas \cite{fisher1969resonance,felsen1994radiation} and emulated in microwave metamaterial circuits \cite{balmain2002resonance}.
Here, we consider the plasmonic equivalent of resonance cone behavior at optical frequencies. 
Below, and in the following sections we show that spontaneous emission from an isolated emitter is enhanced and directed into sub-diffraction resonance cones.
The enhanced emission leads to single photon resonance cones which can be outcoupled to vacuum, thus allowing the study of non-classical light propagation in metamaterials and opening the route to quantum applications.

We consider the case of an emitter such as a quantum dot or dye molecule with a dipole transition, placed within the practical multilayer metal-dielectric metamaterial (Fig.\ref{ResonanceConesFigure}(a)). Preferential emission into the resonance cone is observed for a dipole emitter placed within a 30 nm TiO$_2$ layer surrounded on top and bottom by alternating layers of silver and TiO$_2$. The thickness of the silver layer and TiO$_2$ layers are 10 nm and 30 nm respectively. There are 5 Ag layers and 4  TiO$_2$ layers on each side of the embedded layer.  The resonance cone emission in the finite multilayer realization matches quite closely with the high-$k$ limit predicted by effective medium theory for an ideal hyperbolic metamaterial. The Poynting vector is non-zero only outside the resonance cone ($\theta\geq\theta_{RC}$) and energy tends to flow along the surface of the resonance cone $\theta \approx \theta_{RC}$ (Fig.\ref{ResonanceConesFigure}(c)inset)

%
%
%
\section{Decay Rate Enhancement}

Along with the directional nature of radiation, we show that a large Purcell factor in a broadband range is possible especially when the emitter is embedded inside the metal-dielectric multilayer structure. This enhancement is due to the combination of three effects: a) strong overlap of the plasmonic Bloch modes with the emitter b) presence of a large number of such modes and c) subwavelength confinement of the emitter. The plasmonic Bloch modes of the multilayer structure have a higher momentum than conventional SPPs even away from resonance. This leads to broadband enhancement in the local density of states and these modes are a new radiative decay channel for embedded dye molecules or quantum dots \cite{jacob2012broadband,noginov2010controlling,jacob2010engineering,iorsh2011spontaneous}. This initial research into hyperbolic metamaterials has shown their potential for radiative decay engineering and here we consider the transmission of these modes for applications such as single photon sources.

In the quantum mechanical approach, the decay rate of a single emitter is given by
\begin{equation}
	\Gamma = \frac{2\omega^2}{\hbar c^2}\left(\vec{\mu}\cdot\textrm{Im}\left[\bar{\bar{G}}(r_0,r_0,\omega)\right]\cdot\vec{\mu}\right)
\end{equation}
where $\bar{\bar{G}}(r_0,r_0,\omega)$ is the Green's tensor of the electromagnetic wave-equation \cite{novotny2006principles}, evaluated at the location of the quantum emitter with dipole moment $\vec{\mu}$. Here we use the semi-classical approach developed by Ford and Weber \cite{ford1984electromagnetic} which treats the quantum emitter as a radiating point dipole. In the weak coupling limit, the quantum mechanical approach and the semi-classical theory are equivalent.

Using this approach, we find for a point dipole with dipole moment $\vec{\mu}$ placed a distance $d$ above (or embedded in) a HMM planar slab, the decay rate enhancement (relative to vacuum) to be
\begin{equation}
\beta = \Gamma/\Gamma_0 = (1-\eta) + \eta~\textrm{Re}%
					\left[\int_0^\infty \rho(\lambda,d,\vec{k}) \textrm{d}k_{\parallel}\right]
\label{betaeqn}
\end{equation}

where $k_{\parallel}$ is the wavevector parallel to the interface of the HMM, and $\eta$ is the intrinsic quantum yield.
Here $\rho(\lambda,d,\vec{k})$ is the wavevector resolved local density of states (WLDOS) seen by the emitter, normalized by the LDOS of vacuum.
The WLDOS takes into account the various modes of the system which is captured by the angular reflection spectrum. 
\begin{equation}
\rho(\lambda,d,\vec{k}) = 	\frac{3}{2}\frac{1}{k_1^3}\frac{1}{|\vec{\mu}|^2}%
		   				\frac{k_x}{k_z} \e^{i2k_zd}%
						\left\{%
						\frac{1}{2}\mu^2_\parallel\left[(1+r^{(s)})k_1^2-(1-r^{(p)})k_z^2\right]+\mu_\perp^2(1+r^{(p)})k_{\parallel}^2
						\right\}
\end{equation}
where $r^{(s)}$ and $r^{(p)}$ are the reflection coefficients for $s$- and $p$-polarized light respectively.
$k_1$ is the magnitude of the wavevector in the medium where the dipole resides $k_1 = \sqrt{\epsilon}(\omega/c)$.
Finally $k_z$ is determined from the dispersion relation where the dipole resides $k_z = \sqrt{k_1^2-k_\parallel^2}$.

\begin{figure}
\begin{center}
\includegraphics[width=0.74\textwidth]{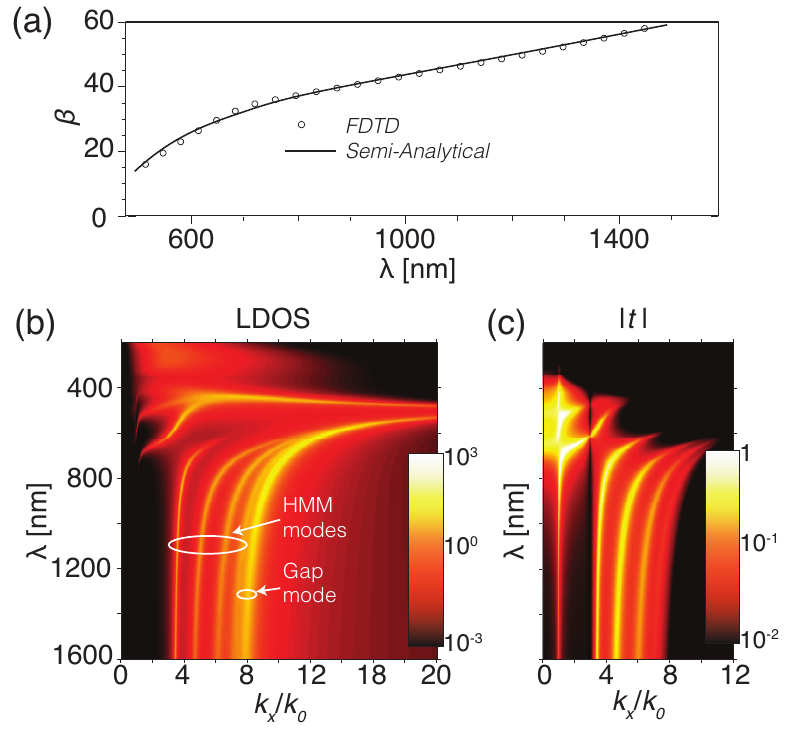}
\caption{(a)  A large decay rate enhancement is predicted across the region of hyperbolic dispersion.  (b)The W-LDOS (normalized by vacuum) available to a quantum emitter embedded in a multilayer realization of a HMM show that emission occurs into bulk waveguide modes of the HMM slab in the region of hyperbolic dispersion. (c) The transmission coefficient (computed using a transfer matrix method for a plane wave launched from the embedded layer) at the top of the multilayer slab shows that the HMM waveguide modes propagate to the top of the HMM slab and can be outcoupled.}
\label{LDOS}
\end{center}
\end{figure}

Figure \ref{LDOS}(a) shows the large total decay rate enhancement $\beta = \Gamma/\Gamma_0$ of an emitter embedded in a  Ag/TiO$_2$ multilayer slab across the region of hyperbolic dispersion ($\lambda>720$ nm) ($\Gamma_0$ is the decay rate in a homogeneous slab of TiO$_2$). The Green's tensor calculation of $\beta$ shows excellent with the full wave FDTD calculation. The monotonically increasing behaviour of $\beta(\lambda)$ is explained by noting that the coupling strength between the emitter and the HMM modes increases at shorter interaction distances, $d/\lambda$. In figure \ref{LDOS}(a) $d$ is fixed, however as $\lambda$ increases the interaction distance decreases; furthermore, material dispersion plays a strong role in the exact nature of the $\beta(\lambda)$ scaling.

%
This enhanced spontaneous emission is channeled to multiple modes all of which carry energy along the above mentioned resonance cones.  We therefore analyze the modal distribution of the spontaneously emitted radiation using the wavevector resolved local density of states (W-LDOS) \cite{cortes2012quantum}. This separates the density of states available to the emitter within the multilayer structure according to the modes characterized by the wavevector parallel to the layers ($k_x$ ).

Figure \ref{LDOS}(b) shows the enhanced W-LDOS available to a single emitter embedded in a 30 nm thick layer of TiO$_2$ clad by realistic finite slabs of Ag/TiO$_2$ multilayer HMM (see figure \ref{ResonanceConesFigure}(a)).
We identify the series of bright vertical bands as high-$k$ HMM waveguide modes that arise from coupled surface-plasmon-polaritons, and identify the last bright line as the gap mode. When the embedded layer thickness is kept above $\approx$ 30 nm, emission into the high-$k$ modes of the HMM and hence resonance cones dominates the total decay enhancement $\beta$. When the embedded layer is smaller, emission into the gap plasmon mode and emission quenching begins to dominate the total decay enhancement $\beta$.

Two severe limitations on plasmonic approaches in general are the finite absorption length in the structure and non-radiative quenching. To study the propagation length of high-$k$ modes in the hyperbolic metamaterial we plot the transfer function of a practical structure.  Figure \ref{LDOS}(c) shows the transmission coefficient at the top of the multilayer HMM slab, and shows that the high-$k$ HMM waveguide modes  propagate a distance of $L \approx 170$ nm to the edge of the slab, and therefore can be out coupled. A detailed account of non-radiative decay and quenching is provided later in this paper. We note that functioning away from resonance and optimizing the embedded layer thickness can lead to a large fraction of spontaneous emission directed into the high-$k$ modes and hence resonance cones.

We emphasize that single photon sources based on the HMM high-$k$ modes are fundamentally different than those based on MIM plasmonic modes \cite{russell2012large,jun2009strong}, slow light modes \cite{yao2009ultrahigh}, localized plasmons \cite{koenderink2009plasmon, chang2006quantum} and slot waveguide modes \cite{quan2009broadband}.
In MIM and slot waveguides, energy flow is along the MIM interfaces; while in HMM systems energy flows through the metal-dielectric multilayers at oblique angles. Furthermore, by examining the band structure in the W-LDOS calculation we observe that the group velocity $v_g = \partial\omega/\partial k$ does not vanish and the HMM waveguide modes are therefore, not slow light modes.

\section{Near field scaling laws}

Peaks in the WLDOS indicate dominant decay channels of the emitter, and as shown below, when the emitter is in the near-field of a HMM slab, it is dominated by unique high-$k$ propagating waveguide modes.
In the following sections, the interaction distance $d$ dependence of $\beta$ and $\rho$ for an effective metamaterial and a physical realizable structure are investigated.
To help elucidate the dependence on the interaction distance we compute the power law dependence of the decay rate enhancement on the distance $n  = \diff{\log{\beta}}{\log{d}}$

\subsection{$\beta(d)$ for Effective HMM Slabs}

\begin{figure}[ht]
\begin{center}
	\includegraphics[width=0.9\textwidth]{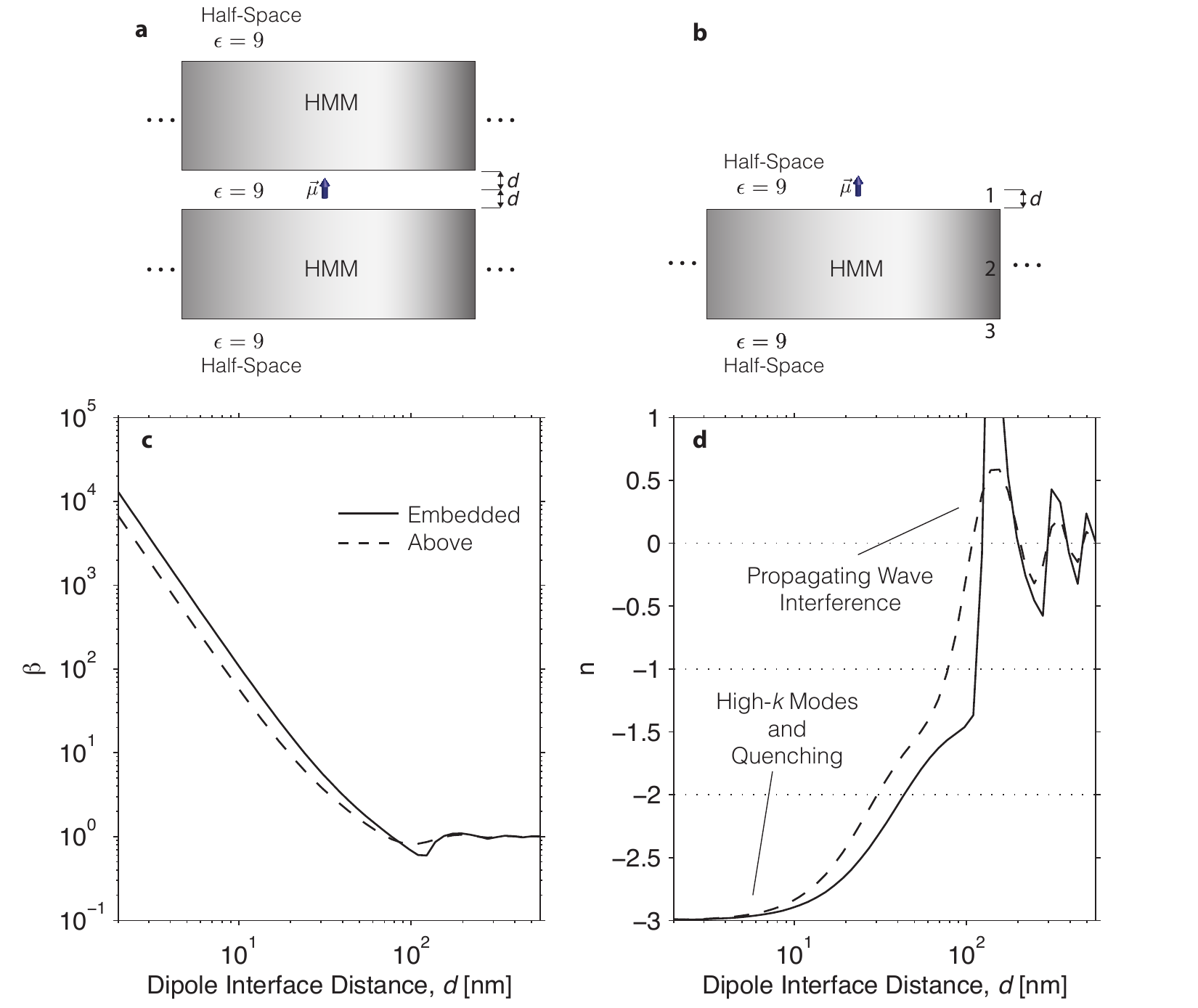}
	\caption{{\bf (a)} Schematic of a dipole embedded symmetrically between two 370 nm thick HMM slabs composed of Ag/TiO$_2$ with a silver filling fraction of 0.25. Type 2 hyperbolic dispersion is predicted above $\lambda \approx 720$ nm. The coupling to the metamaterial changes as the embedded layer thickness is varied. {\bf (b)} Schematic of a dipole placed above one of the finite HMM slabs from {\bf a}. 
	({\bf c}) Decay rate enhancement $\beta$ at $\lambda=900$ nm is stronger when the dipole is embedded between two HMMs. 
	{\bf (d)} In the extreme near-field, the decay rate enhancement varies as the inverse cube of the interaction distance which is attributed to emission into high-$k$ HMM modes and quenching. In the far-field, the decay rate enhancement is governed by plane wave interference effects.
	}
	\label{EMTSWEEP}
\end{center}
\end{figure}

In this section, we treat the HMMs using effective medium theory.
The systems studied are shown schematically in figure \ref{EMTSWEEP}(a,b).
To compute the reflection coefficients required to calculate $\beta$ and $\rho$ for each of these two geometries, the fresnel reflection coefficients for plane waves incident on a finite slab are used. For the case of a dipole embedded between two finite HMM slabs the reflection coefficients are computed using plane wave reflection coefficients in the {\it tunnel junction geometry} which can found in \cite{ford1984electromagnetic}.


The calculated $\beta$ and $n$ are shown in figure \ref{EMTSWEEP}(c,d) for a {\it vertically oriented dipole} at a wavelength of $\lambda = 900$ nm and for 370 nm thick metamaterial slabs consisting of Ag/TiO$_2$ with a metal filling fraction of 0.25.
Realistic loss and dispersion were used for the effective medium computation \cite{JohnsonAndChristy,NanoHub}.
The metamaterial is predicted to be type 2 ($\epsilon_\parallel<0$, $\epsilon_\perp>0$) above $\lambda \approx 720$ nm; the predicted permittivity dispersion of this structure are shown in figure \ref{EMT}a.
At $\lambda = 900$ nm, the dielectric response of the parallel and perpedicular compenent of the dielectric tensor are $\epsilon_\parallel = -4.3 +i0.4$ and $\epsilon_\perp = 11.0+i0.07$

Due to confinement effects, the embedded dipole has a larger enhanced decay rate than the dipole simply above the effective medium HMM slab.
Furthermore, in the limit of the dipole located very close to the HMM interface(s) $d<<\lambda/\sqrt{\epsilon}$, the decay rate enhancement is proportional to the inverse cube of the interaction distance $\beta\sim d^{-3}$.
This extreme near-field behaviour is attributed to emission into the unique high-$k$ modes of the HMM and emission into lossy surface waves (quenching) which is discussed in more detail in the following sections.
For large interaction distances where $d\sim\lambda/\sqrt{\epsilon}$, plane wave interference/field intensity effects govern the decay rate enhancement; this is discussed in more detail in the following sections.

\subsection{$\beta(d)$ for Practical Realization of HMMs: Multilayer Slabs}

\begin{figure}[ht]
\begin{center}
	\includegraphics[width=0.9\textwidth]{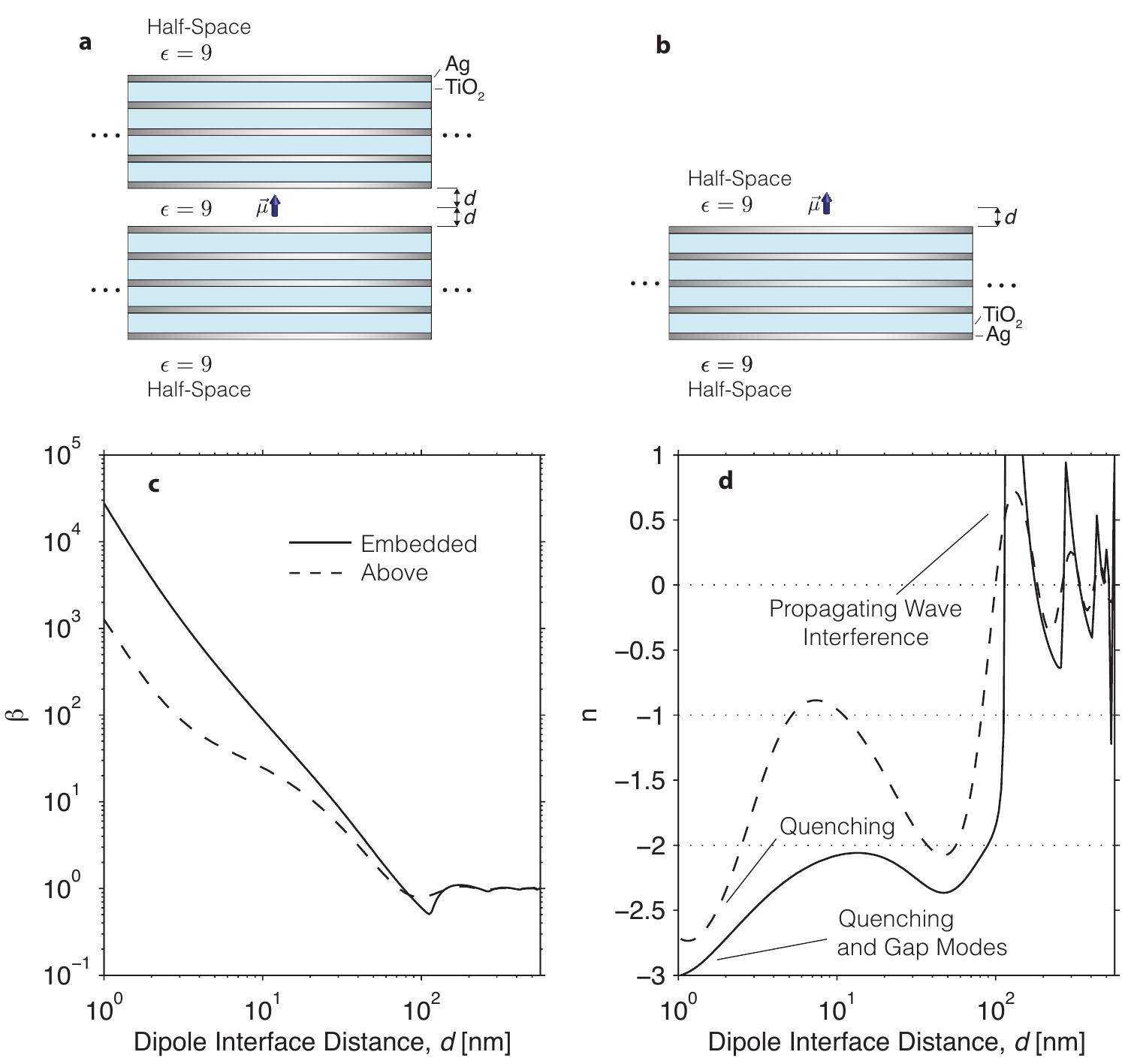}
	\caption{{\bf (a)} Schematic of a dipole embedded symmetrically between two 4.5 period multilayer HMM slabs composed of Ag/TiO$_2$ with thicknesses 10/30 nm respectively. Type 2 hyperbolic dispersion is predicted above $\lambda \approx 720$ nm. The embedded layer thickness is varied. {\bf (b)} A dipole is placed above one of the finite HMM slabs from {\bf a}. The dipole/interface distance is varied.
	({\bf c}) Decay rate $\beta$ calculations at $\lambda=900$ nm show that the enhancement is considerably stronger when the dipole is embedded between two multilayer HMMs. 
	{\bf (d)} In the extreme near-field, the decay rate enhancement varies as the inverse cube of the interaction distance; this is attributed to (\emph{i}) emission into a gap plasmon mode and quenching for the embedded emitter and (\emph{ii}) emission quenching when the emitter is placed above the multilayer.
	In the far-field, the decay rate enhancement is governed by propagating wave interference effects.
	}
	\label{MLSWEEP}
\end{center}
\end{figure}

In this section, we treat the HMMs using a physical realizable multilayer structure.
The systems studied are shown schematically in figure \ref{MLSWEEP}(a,b).
To compute the reflection coefficients required to calculate $\beta$ and $\rho$, a transfer matrix method is used. The transfer matrix method can be generalized in the tunnel junction geometry, where the dipole is embedded between two multilayer stacks.

%
%
%
%
%
%
%
%
%

The calculated $\beta$ and $n$ are shown in figure \ref{MLSWEEP}(c,d) at a wavelength of $\lambda = 900$ nm and a 4.5 period Ag/TiO$_2$ multilayer with thicknesses 10/30 nm respectively.
Realistic loss and dispersion were used for the calculations \cite{JohnsonAndChristy,NanoHub}.
At $\lambda = 900$ nm, TiO$_2$ is lossless and the loss in the structure is due to silver which has $\epsilon_{Ag} \approx -40 + i1.7$
The metamaterial is predicted to be type 2 ($\epsilon_\parallel<0$, $\epsilon_\perp>0$) at this wavelength; the predicted effective medium dispersion of this structure can be found in figure \ref{EMT}(a).

We observe that in general, the embedded dipole has a larger enhanced decay rate than the dipole simply above the multilayer.
For large interaction distances where $d\sim\lambda/\sqrt{\epsilon}$, plane wave interference/field intensity effects govern the decay rate enhancement; this is observed in the low-$k$ LDOS shown in following sections.
When the dipole located extremely close to the HMM interface $d<<\lambda/\sqrt{\epsilon}$, the behaviour of the dipole located above the multilayer and the dipole embedded between the multilayers is different: there are different modes available for dipole to decay into in each geometry.
This is elucidated through the high-$k$ LDOS discussed in the following sections.

\section{Quenching}

The wave vector resolved density of states $\rho(\lambda,d,\vec{k})$ provides insight into the dominate decay channels of the emitter and helps explain the dependence of $\beta$ on the interaction distance.
Here we show the preferred decay channels of a dipole at various distances from an effective medium HMM and a physically realizable multilayer HMM.

\subsection{Far-Field Interference}

\begin{figure}[ht]
\begin{center}
	\includegraphics[width=0.95\textwidth]{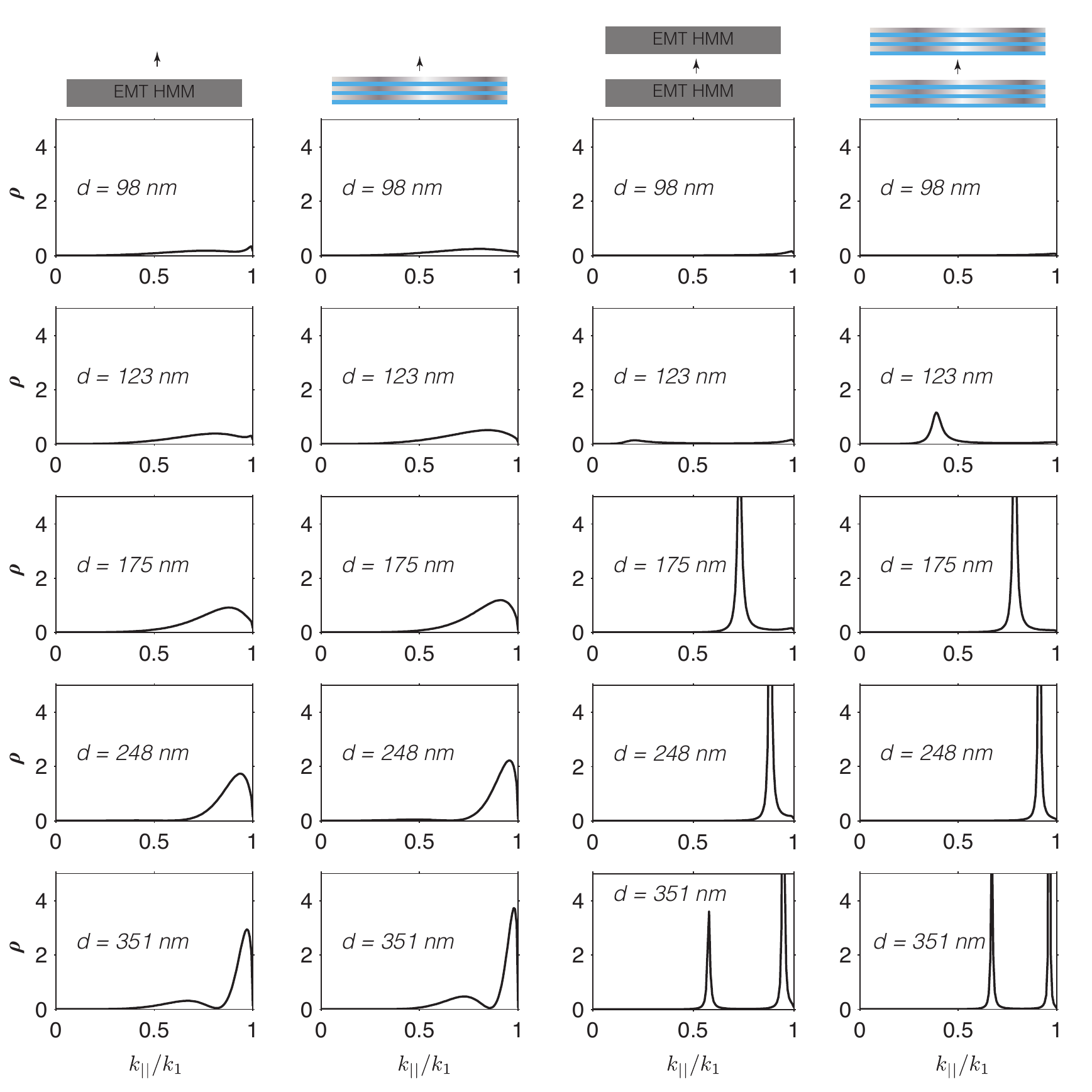}
	\caption{Far-field LDOS: Calculated wavevector resolved LDOS at various far-field interaction distances at $\lambda = 900$ nm show that when $d\geq1/2~\lambda/\sqrt{\epsilon}\approx 150$ nm, low-$k$ interactions dominate. As the dipole is brought closer (embedded layer shrunk), emission is shifted into high-$k$ modes. The peaks in the low-$k$ LDOS are due to propagating wave interference.}
	\label{lowk}
\end{center}
\end{figure}

Figure \ref{lowk} shows the calculated low-$k$ LDOS for the four geometries considered in the previous two sections.
We see distinct peaks in the LDOS which can attributed to propagating wave interference effects.
When the waves interfere constructively at the location of the emitter, there is an enhancement in the electric field intensity, and thus an enhancement in the decay rate of the emitter.
In the embedded case, the propagating wave interference is due in Fabry-Perot modes.
Purely dielectric microresonators used to enhance spontaneous emission utilize these low-$k$ modes.

As the dipole is brought closer to the metal layers (embedded layer shrunk), the coupling to the high-$k$ modes increases.For interaction distances smaller than approximately half the wavelength in the medium $d \leq 1/2~\lambda/\sqrt{\epsilon} \approx 150$ nm near-field interactions begin to dominate, and emission into  propagating waves (low-$k$ modes) no longer takes place.

\subsection{ Near-Field Modes: EMT}

\begin{figure}[ht]
\begin{center}
	\includegraphics{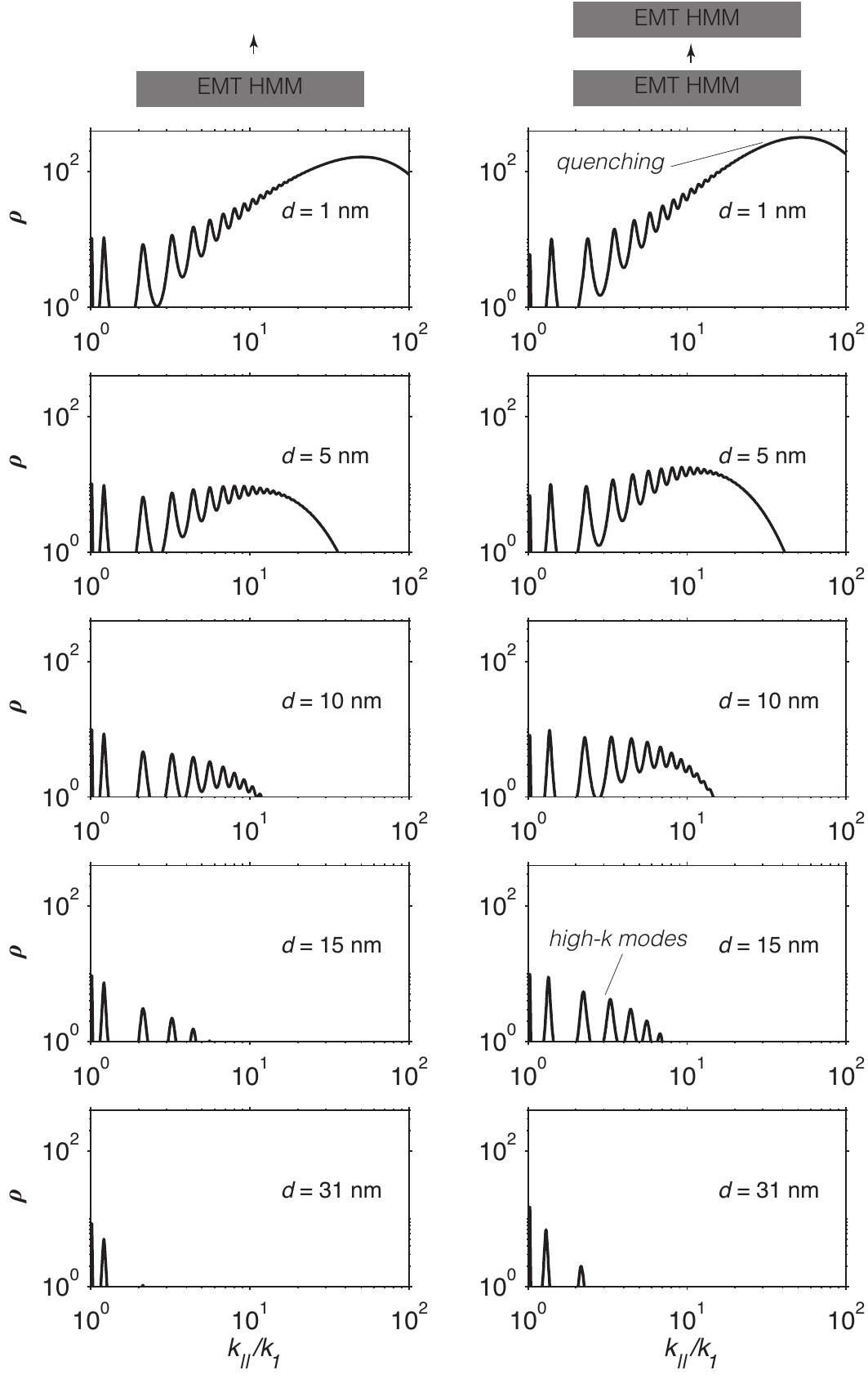}
	\caption{Near-field LDOS - Effective Medium Theory: Calculated wave vector resolved LDOS at various near-field interaction distances at $\lambda = 900$ nm show that as the interaction distance $d$ is decreased, additional high-$k$ modes become accessible to the emitter. The HMM modes are recognizable as sharp peaks in $\rho$ and are attributed to propagating waveguide modes of the HMM. When the emitter is embedded between two HMM slabs the coupling to high-$k$ HMM modes is slightly stronger and the onset of the strong near-field interactions occurs at larger $d$. At very small $d$, emission quenching begins to dominate. This is recognized as the wide smooth peak in $\rho$. Emission into the high-$k$ HMM modes and emission quenching cause the total decay rate enhancement $\beta$ to scale as $d^{-3}$ in the extreme near-field.}
	\label{HIGHKEMT}
\end{center}
\end{figure}

When the dipole is located in the near field of the metamaterial, emission is directed into the high-$k$ modes of HMM slab.
In the effective medium limit, there is no upper cut-off on the allowed wave vectors, and as the dipole is brought closer to (embedded layer shrunk), extremely larger wave vector modes become available to the emitter.
At very small distances, $d/(\lambda/\sqrt{\epsilon}) \ll 1$ emission into these infinite amount of high-$k$ propagating modes of the HMM contribute to the enhanced total decay rate as the inverse cube of the interaction distance $\beta\sim1/d^3$ \cite{krishnamoorthy2012topological}.
However, there is a competing decay channel, namely quenching.
If the emitter is brought extremely close to the interface then emission into lossy surface waves begins to dominate the emission into high-$k$ HMM modes, thus $\beta$ is mainly due to non-radiative decay.
This is demonstrated in figure \ref{HIGHKEMT} which shows the WLDOS $\rho(\lambda,d,\vec{k})$ for a perpendicular oriented dipole placed above a HMM slab and embedded between two HMM slabs.
The geometry is the same as that used in figure \ref{EMTSWEEP}.
As the emitter is brought closer to the HMM slab (embedded layer shrunk) coupling occurs to increasingly larger high-$k$ propagating modes.
These modes are interpreted as confined waveguide modes of the HMM slab.
In the LDOS, these modes are manifested as distinct, relatively narrow peaks.
When the emitter is at extreme near-field distances quenching begins to dominate the total decay enhancement $\beta$.
Quenching (emission into lossy surface waves) appears as a broad peak in the LDOS.

By separating the LDOS integral which determines the decay rate enhancement (equation \ref{betaeqn}) into the two dominant decay channels, the decay rate enhancement due to the HMM waveguide modes (HMM) and the lossy surface waves (LSW) can be estimated from $\beta \approx \beta_{HMM}+\beta_{LSW}$. Such a separation is possible by calculating the area under the WLDOS corresponding only to HMM modes or to LSW.
By examining  figure \ref{HIGHKEMT} we conclude that at optical wavelengths and at intermediate near-field distance $d\approx5-20$ nm, emission into high-$k$ HMM waveguide modes and hence resonance cones dominates in the EMT limit.
For interaction distances of $d<5$ nm, the emission is quenched, and the decay enhancement is largely unusable.

\subsection{ Near-Field Modes: Multilayer Realization}

\begin{figure}[ht]
\begin{center}
	\includegraphics{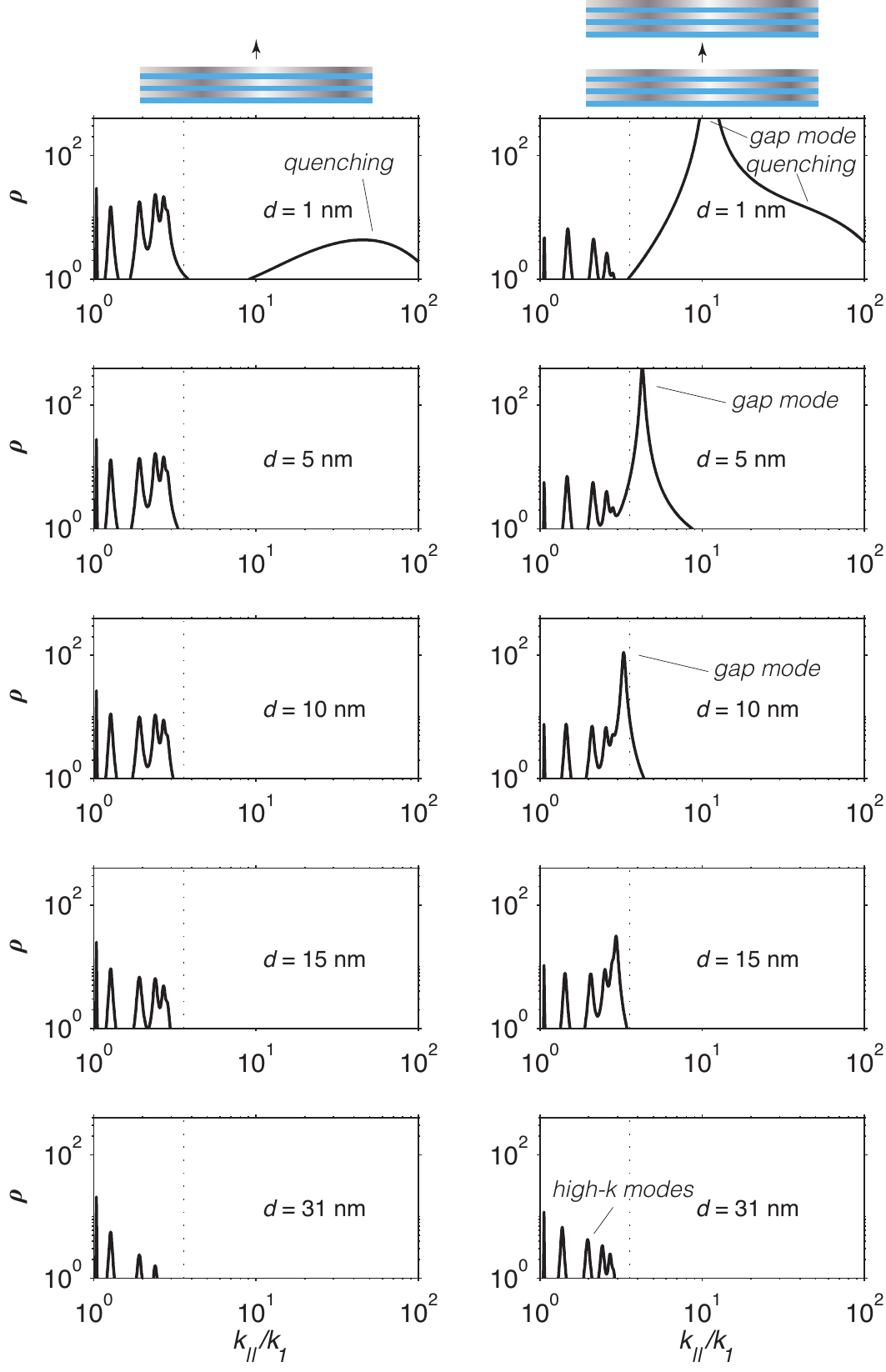}
	\caption{ Near-field LDOS - Multilayer System: 
	Calculated wave vector resolved LDOS at various interaction distances at $\lambda = 900$ nm show that as the interaction distance $d$ is decreased, additional high-$k$ modes become accessible to the emitter. No high-$k$ HMM modes exist above the cut-off (dashed line) related to the unit cell size. In the case of the emitter above the multilayer slab, the onset of emission into lossy surface waves (smooth broad peak) occurs at much smaller distances than in the EMT case. In the embedded case, we also see the presence of a gap plasmon mode, identified by the large peak which shifts with embedded layer thickness. This emission can be suppressed using the appropriate gap size. Again, the onset of emission into lossy surface waves occurs at smaller distances than in the EMT case.}
	\label{HIGHKML}
\end{center}
\end{figure}

In the effective medium limit there is no upper cut-off on the transverse wave vector allowed by the medium.
In a physical realizable system, the maximum transverse wave vector $k_\parallel$ that can be supported is inversely proportional to the unit cell size $1/a$.
As a result, there is a finite number of high-$k$ modes supported by the multilayer HMM structure.
This is demonstrated in figure \ref{HIGHKML} which shows the LDOS dependence on the interaction distances $d$ for the two geometries shown in figure \ref{MLSWEEP}.

As the emitter is brought closer the multilayer slab (embedded layer shrunk), additional high-$k$ modes of the HMM appear to the emitter.
We observe the presence of the {\it finite period cut-off}, above which no high-$k$ HMM waveguide modes are present.
The onset of quenching (broad smooth peak in LDOS) occurs at much closer distances in the physical multilayer system than in the EMT approximation (compare figures \ref{HIGHKEMT} and \ref{HIGHKML}).
In the case of an emitter embedded between two multilayer HMMs, we see the presence of a gap plasmon mode (metal-insulator-metal plasmon mode) when the embedded layer thickness (gap) is smaller than about 15 nm.
The gap mode can identified by a large sharp peak in the LDOS.

Similar to above, the LDOS integral which determines the decay rate enhancement $\beta$ (equation \ref{betaeqn}) is separated into the three dominant decay channels: high-$k$ bloch plasmon modes (HMM), the gap plasmon mode (GPM), and lossy surface waves (LSW).
The decay rate enhancement from each can then be determined from $ \beta \approx \beta_{HMM}+ \beta_{GPM} + \beta_{LSW}$
Utilizing  figure \ref{HIGHKML}, we conclude that at optical wavelengths and at intermediate near-field interaction distances of $d\approx10-20$ nm,  emission into high-k modes ($\beta_{HMM}$) is much greater than the other two contributions.
These high-$k$ modes contribute to emission into highly directive resonance cones.
By moving the emitter closer to the multilayer HMM, emission into the gap mode and quenching begins to dominate the total decay enhancement.
However, in contrast to the EMT case, when the emitter is embedded in an extremely small layer, emission into the gap plasmon mode dominates, not quenching, the other decay channels.
This result is in agreement with \cite{russell2012large}.
The gap plasmon mode does not contribute to emission into resonance cones.

\section{Topological transition assisted pumping scheme}

We now address the critical challenge of pumping embedded emitters in metamaterials. Quite often the large impedance mismatch with vacuum leads to inefficient coupling of radiation to emitters placed inside such structures. We design our metamaterial to have a transparency window at the pumping wavelength and a large density of states (plasmonic response) only in the wavelength of emission.

\begin{figure}[ht]
\begin{center}
\includegraphics[width=0.74\textwidth]{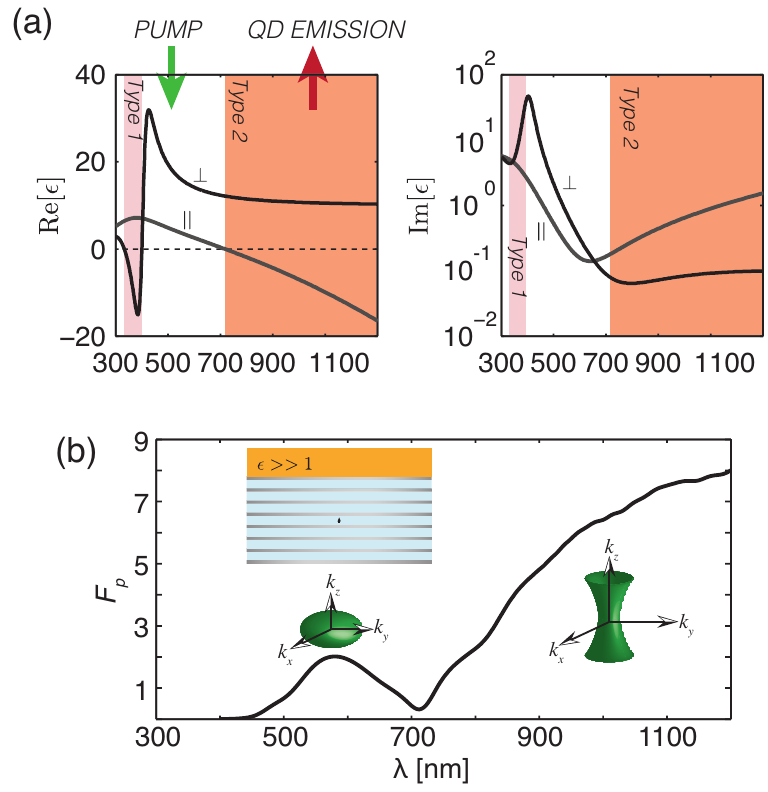}
\caption{(a) Parallel and perpendicular components of the dielectric permittivity predicted by EMT for a 10 nm and 30 nm Ag/TiO$_2$ multilayer structure. A topological transition between elliptical dispersion and hyperbolic dispersion occurs just above $\lambda_{OTT} \approx 720$ nm. (b) Using a large dielectric constant half-space superstrate ($\epsilon >> 1$), the high-$k$ modes of the resonance cones are outcoupled into well defined propagating modes in the far-field of the dielectric over a broadband spectral range. There is excellent agreement between the topological transition predicted by EMT and the large Purcell factor achieved in the practical multilayer structure.}
\label{EMT}
\end{center}
\end{figure}

Figure \ref{EMT}(a) shows the effective medium dielectric response of the Ag/TiO$_2$ multilayer. Below $\lambda_{OTT} \approx 720$ nm the multilayer slab is effectively a dielectric ($\epsilon_\parallel > 0, \epsilon_\perp > 0 $) and transparent to normal incident light. Above $\lambda_{OTT} \approx 720$ nm the multilayer slab is a type 2 HMM ($\epsilon_\parallel > 0, \epsilon_\perp < 0 $), where enormous increase in the LDOS yields large decay rate enhancement $\beta$ into high-$k$ modes of the HMM. This significantly different behaviour is attributed to an optical topological transition (OTT) in the iso-frequency ($k$-space) surface from a closed ellipsoid to an open hyperboloid at $\lambda_{OTT} \approx 720$ nm which drastically changes the decay channels available to the emitter (see Eq.\ref{dispersion}) \cite{krishnamoorthy2012topological}. The difference in the electromagnetic responses above and below the topological transition allows for a novel realistic pumping scheme: a single emitter can be pumped where the HMM is transparent, and in turn have its lower energy emission enhanced and extracted in the type 2 hyperbolic dispersion region.



To understand the nature of spontaneously emitted radiation we consider the case of a high index superstrate ($\epsilon >> 1$) and define the Purcell factor using only propagating waves that reach the far-field.  As expected, without the superstrate all the modes are confined to the structure and do not out couple to the far-field. However a very interesting transitional behavior is seen in the far-field with the high index superstrate. Figure \ref{EMT}b shows the predicted far-field Purcell factor $F_p$ of the emitter inside a practical metal-dielectric multilayer with a high-index superstrate ($\epsilon \approx 30$). Enhanced spontaneous emission reaches the far-field of the superstrate and there is a large increase in the Purcell factor exactly at the transition wavelength predicted by effective medium theory ($\lambda_{OTT}$).  Note the broad bandwidth in which this Purcell factor occurs.

\section{Outcoupling}

\begin{figure}[ht]
\begin{center}
\includegraphics[width=0.74\textwidth]{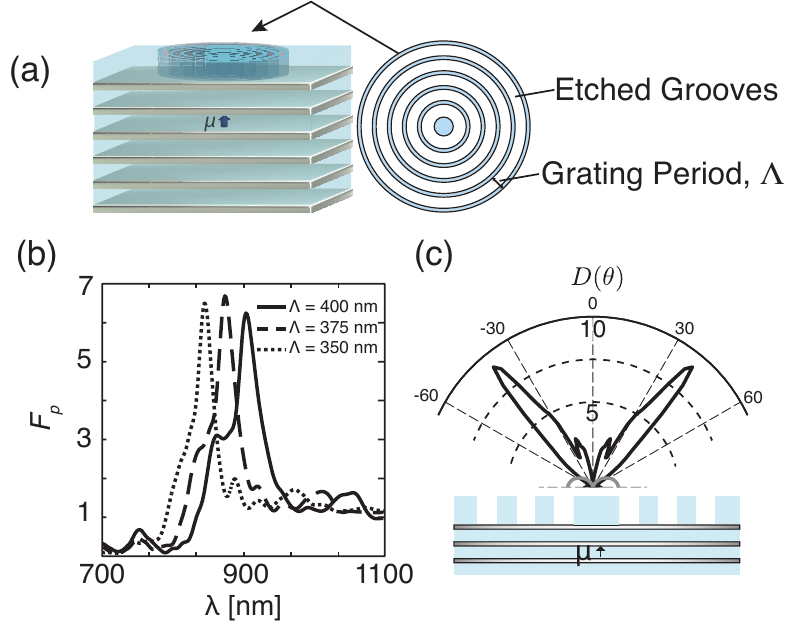}
\caption{(a) A subwavelength cylindrical {\it Bullseye} grating can be used to scatter the high-$k$ states which are intrinsically confined to the HMM slab, into vacuum propagating modes. (b) FDTD simulations show that the spectral location of the maximum far-field Purcell factor $F_p$ can be tuned by varying the grating period. (c) The directivity $D = P(\theta)/(P_{tot}/4\pi)$ is shown at the location of maximum far-field enhancement ($\Lambda = 400$ nm; $\lambda_{max} \approx 900$ nm). The emission from the bullseye occurs into a highly directive conical shell. The small grey curve shows the directivity of a dipole in TiO$_2$. Directivity is a measure of the power density directed along a particular direction relative to an isotropic point source emitting the same total power.}
\label{Grating}
\end{center}
\end{figure}

We now address the outcoupling of high-$k$ modes which has been the significant limiting factor for all applications of hyperbolic metamaterials \cite{krishnamoorthy2012topological,noginov2010controlling}. Given the cylindrical symmetry of the dipole/multilayer system, we propose a Bullseye grating structure \cite{lezec2002beaming} to out couple the resonance cone to vacuum.
Figure \ref{Grating}(a) shows the proposed bullseye structure etched into an 85 nm TiO$_2$ layer on top of the Ag/TiO$_2$ HMM structure.
Figure \ref{Grating}(b) shows that the far-field Purcell factor can exceed 6, and the spectral location of the maximum $F_p$ can be tuned by varying the Bullseye grating period $\Lambda$.
It is worth emphasizing that the colour selective out coupling by the grating is a resonant phenomenon \cite{de2012tailoring}, however the magnitude of the $F_p$ is due to the decay rate enhancement into the high-$k$ HMM modes (resonance cones), which is fundamentally non-resonant and broadband.

Since the quantum emitter couples into specific metamaterial modes with a unique lateral wave vector $k_x$ (figure \ref{LDOS}a), the Bullseye grating scatters these modes into well defined free space modes. The directivity $D$ gives a measure of the enhanced directivity of a given emission pattern relative to an isotropic point source; it is defined as $D(\theta) = P(\theta)/(P_{rad}/4\pi)$ where $P(\theta)$ is the power density at a given far-field angle $\theta$ and $P_{rad}/4\pi$ is the total power radiated into the far-field averaged over all solid angles. Figure \ref{Grating}(c) shows the directivity on a linear scale for a grating period of $\Lambda = 400$ nm at the wavelength of maximum far-field Purcell factor $\lambda\approx 900$ nm.
We see that the Bullseye scatters the resonance cone into a highly directive well defined spatial mode: a thin conical shell with a half angle of $\theta \approx 34^o$.

In this investigation, we have considered a single period Bullseye grating which can be used to outcouple a single wavelength despite the fact that the HMM structure provides a spectrally broad decay rate enhancement.
Recent work on broadband plasmonic scattering \cite{ChirpedGratings,ChirpedGratings2} has demonstrated that efficient outcoupling of plasmons across the optical spectrum can be achieved with chirped gratings. Modification of the Bullseye grating design can lead to broadband outcoupling of the single photons. The HMM approach presented in this paper can be far more advantageous than microcavities for coupling with broadband emitters such as NV centers in diamond.

\section{Conclusion}

In conclusion, we have proposed a device using plasmonic metamaterials to efficiently extract single photons from broadband quantum emitters such as NV centers in diamond. Detailed analysis of quenching and non-radiative decay  shows that the device can achieve a broadband purcell effect and directional far-field spontaneous emission despite the metal losses. Low mode volume confinement of the emitter and the ability to function away from plasmonic resonances holds the key to efficient outcoupling of light. The proposed structure will lead to a platform for understanding propagation of non-classical light in plasmonic metamaterials and can be integrated with other plasmonic structures for future nanoscale quantum information applications.

\section*{{\footnotesize Acknowledgements}}
Z.Jacob acknowledges E. E. Narimanov and V. M. Shalaev for discussions. This work was supported by the Natural Science and Engineering Research Council of Canada (NSERC), Alberta Innovates Technology Futures (AITF),  Alberta Nanobridge and Canadian School of Energy and Environment.


\begin{thebibliography}{10}

\bibitem{nielsen2010quantum}
Michael~A. Nielsen and Isaac~L. Chuang.
\newblock {\em Quantum Computation and Quantum Information}.
\newblock Cambridge University Press, December 2010.

\bibitem{altewischer2002plasmonassisted}
E.~Altewischer, {MP} Van~Exter, and {JP} Woerdman.
\newblock Plasmon-assisted transmission of entangled photons.
\newblock {\em Nature}, 418(6895):304--306, 2002.

\bibitem{huck2009demonstration}
A.~Huck, S.~Smolka, P.~Lodahl, {A.S.} {SU00F8rensen}, A.~Boltasseva,
  J.~Janousek, and {U.L.} Andersen.
\newblock Demonstration of quadrature-squeezed surface plasmons in a gold
  waveguide.
\newblock {\em Physical review letters}, 102(24), 2009.

\bibitem{dimartino2012quantum}
Giuliana Di~Martino, Yannick Sonnefraud, Stephane Kena-Cohen, Mark Tame,
  Aahin~K. Azdemir, M.~S. Kim, and Stefan~A. Maier.
\newblock Quantum statistics of surface plasmon polaritons in metallic stripe
  waveguides.
\newblock {\em Nano Letters}, 12(5):2504--2508, May 2012.

\bibitem{jacob2011plasmonics}
Z.~Jacob and V.~M. Shalaev.
\newblock Plasmonics goes quantum.
\newblock {\em Science}, 334(6055):463--464, 2011.


\bibitem{jacob2012quantum}
Z.~Jacob
\newblock Quantum Plasmonics
\newblock {\em MRS Bulletin}, 37(8),761--767,2012



\bibitem{lounis2005singlephoton}
B.~Lounis and M.~Orrit.
\newblock Single-photon sources.
\newblock {\em Reports on Progress in Physics}, 68(5):1129, 2005.

\bibitem{grangier2004focus}
P.~Grangier, B.~Sanders, and J.~Vuckovic.
\newblock Focus on single photons on demand.
\newblock {\em New Journal of Physics}, 6(1), 2004.

\bibitem{aharonovich2011diamond}
I.~Aharonovich, {A.D.} Greentree, and S.~Prawer.
\newblock Diamond photonics.
\newblock {\em Nature Photonics}, 5(7):397--405, 2011.

\bibitem{babinec2010diamond}
T.~M. Babinec, B.~J.~M. Hausmann, M.~Khan, Y.~Zhang, J.~R. Maze, P.~R. Hemmer,
  and M.~Loncar.
\newblock A diamond nanowire single-photon source.
\newblock {\em Nature nanotechnology}, 5(3):195--199, 2010.

\bibitem{choy2011enhanced}
J.T. Choy, B.J.M. Hausmann, T.M. Babinec, I.~Bulu, M.~Khan, P.~Maletinsky,
  A.~Yacoby, and M.~Loncar.
\newblock Enhanced single-photon emission from a diamond-silver aperture.
\newblock {\em Nature Photonics}, 5(12):738--743, 2011.

\bibitem{huck2011controlled}
A.~Huck, S.~Kumar, A.~Shakoor, and {U.L.} Andersen.
\newblock Controlled coupling of a single nitrogen-vacancy center to a silver
  nanowire.
\newblock {\em Physical Review Letters}, 106(9):96801, 2011.

\bibitem{tanaka2010multifold}
K.~Tanaka, E.~Plum, J.~Y. Ou, T.~Uchino, and N.~I. Zheludev.
\newblock Multifold enhancement of quantum dot luminescence in plasmonic
  metamaterials.
\newblock {\em Physical review letters}, 105(22):227403, 2010.

\bibitem{ford1984electromagnetic}
GW~Ford and WH~Weber.
\newblock Electromagnetic interactions of molecules with metal surfaces.
\newblock {\em Physics Reports}, 113(4):195--287, 1984.

\bibitem{sun2007practicable}
G.~Sun, J.B. Khurgin, and R.A. Soref.
\newblock Practicable enhancement of spontaneous emission using surface
  plasmons.
\newblock {\em Applied physics letters}, 90(11):111107--111107, 2007.

\bibitem{esteban2010optical}
R.~Esteban, T.~V. Teperik, and J.~J. Greffet.
\newblock Optical patch antennas for single photon emission using surface
  plasmon resonances.
\newblock {\em Physical review letters}, 104(2):26802, 2010.

\bibitem{friedler2009solidstate}
I.~Friedler, C.~Sauvan, J.~P. Hugonin, P.~Lalanne, J.~Claudon, and J.~M.
  Gérard.
\newblock Solid-state single photon sources: the nanowire antenna.
\newblock {\em Optics express}, 17(4):2095--2110, 2009.

\bibitem{lee2011planar}
K.~G. Lee, X.~W. Chen, H.~Eghlidi, P.~Kukura, R.~Lettow, A.~Renn,
  V.~Sandoghdar, and S.~Götzinger.
\newblock A planar dielectric antenna for directional single-photon emission
  and near-unity collection efficiency.
\newblock {\em Nature Photonics}, 5(3):166--169, 2011.

\bibitem{englund2010deterministic}
D.~Englund, B.~Shields, K.~Rivoire, F.~Hatami, J.~Vu{\v{c}}kovi{\'c}, H.~Park,
  and MD~Lukin.
\newblock Deterministic coupling of a single nitrogen vacancy center to a
  photonic crystal cavity.
\newblock {\em Nano letters}, 10(10):3922, 2010.

\bibitem{xiong2007twodimensional}
Y.~Xiong, Z.~Liu, C.~Sun, and X.~Zhang.
\newblock Two-dimensional imaging by far-field superlens at visible
  wavelengths.
\newblock {\em Nano letters}, 7(11):3360--3365, 2007.

\bibitem{kabashin2009plasmonic}
{AV} Kabashin, P.~Evans, S.~Pastkovsky, W.~Hendren, {GA} Wurtz, R.~Atkinson,
  R.~Pollard, {VA} Podolskiy, and {AV} Zayats.
\newblock Plasmonic nanorod metamaterials for biosensing.
\newblock {\em Nature materials}, 8(11):867--871, 2009.

\bibitem{elser2007nonlocal}
J.~Elser, V.A. Podolskiy, I.~Salakhutdinov, and I.~Avrutsky.
\newblock Nonlocal effects in effective-medium response of nanolayered
  metamaterials.
\newblock {\em Applied physics letters}, 90(19):191109--191109, 2007.

\bibitem{smith2004partial}
D.~R. Smith, D.~Schurig, J.~J. Mock, P.~Kolinko, and P.~Rye.
\newblock Partial focusing of radiation by a slab of indefinite media.
\newblock {\em Applied physics letters}, 84(13):2244--2246, 2004.

\bibitem{fisher1969resonance}
R.~K. Fisher and R.~W. Gould.
\newblock Resonance cones in the field pattern of a short antenna in an
  anisotropic plasma.
\newblock {\em Physical Review Letters}, 22(21):1093--1095, 1969.

\bibitem{felsen1994radiation}
L.B. Felsen, N.~Marcuvitz, et~al.
\newblock {\em Radiation and scattering of waves}.
\newblock IEEE press Piscataway, NJ, 1994.

\bibitem{balmain2002resonance}
K.~G. Balmain, A.~A.~E. Luttgen, and P.~C. Kremer.
\newblock Resonance cone formation, reflection, refraction, and focusing in a
  planar anisotropic metamaterial.
\newblock {\em Antennas and Wireless Propagation Letters, {IEEE}},
  1(1):146--149, 2002.

\bibitem{jacob2012broadband}
Zubin Jacob, Igor~I Smolyaninov, and Evgenii~E Narimanov.
\newblock Broadband purcell effect: Radiative decay engineering with
  metamaterials.
\newblock {\em Applied Physics Letters}, 100(18):181105--181105--4, May 2012.

\bibitem{noginov2010controlling}
M.~A. Noginov, H.~Li, Y.~A. Barnakov, D.~Dryden, G.~Nataraj, G.~Zhu, C.~E.
  Bonner, M.~Mayy, Z.~Jacob, and E.~E. Narimanov.
\newblock Controlling spontaneous emission with metamaterials.
\newblock {\em Optics letters}, 35(11):1863--1865, 2010.

\bibitem{jacob2010engineering}
Z.~Jacob, J.Y. Kim, GV~Naik, A.~Boltasseva, EE~Narimanov, and VM~Shalaev.
\newblock Engineering photonic density of states using metamaterials.
\newblock {\em Applied Physics B: Lasers and Optics}, 100(1):215--218, 2010.

\bibitem{iorsh2011spontaneous}
I.~Iorsh, A.~Poddubny, A.~Orlov, P.~Belov, and Y.~S. Kivshar.
\newblock Spontaneous emission enhancement in metal-dielectric metamaterials.
\newblock {\em Physics Letters A}, 2011.

\bibitem{novotny2006principles}
L.~Novotny and B.~Hecht.
\newblock {\em Principles of nano-optics}.
\newblock Cambridge Univ. Press, 2006.

\bibitem{cortes2012quantum}
C.~L. Cortes, W.~Newman, S.~Molesky, and Z.~Jacob.
\newblock Quantum nanophotonics using hyperbolic metamaterials.
\newblock {\em Journal of Optics}, 14(6):063001, 2012.

\bibitem{russell2012large}
K.~J. Russell, T.~L. Liu, S.~Cui, and E.~L. Hu.
\newblock Large spontaneous emission enhancement in plasmonic nanocavities.
\newblock {\em Nature Photonics}, 6(7):459--462, 2012.

\bibitem{jun2009strong}
Y.~C. Jun, R.~Pala, and M.~L. Brongersma.
\newblock Strong modification of quantum dot spontaneous emission via gap
  plasmon coupling in metal nanoslits†.
\newblock {\em The Journal of Physical Chemistry C}, 114(16):7269--7273, 2009.

\bibitem{yao2009ultrahigh}
P.~Yao, C.~Van~Vlack, A.~Reza, M.~Patterson, M.~M. Dignam, and S.~Hughes.
\newblock Ultrahigh purcell factors and lamb shifts in slow-light metamaterial
  waveguides.
\newblock {\em Physical Review B}, 80(19):195106, 2009.

\bibitem{koenderink2009plasmon}
A.~F. Koenderink.
\newblock Plasmon nanoparticle array waveguides for single photon and single
  plasmon sources.
\newblock {\em Nano letters}, 9(12):4228--4233, 2009.

\bibitem{chang2006quantum}
D.~E. Chang, A.~S. Sørensen, P.~R. Hemmer, and M.~D. Lukin.
\newblock Quantum optics with surface plasmons.
\newblock {\em Physical review letters}, 97(5):53002, 2006.

\bibitem{quan2009broadband}
Q.~Quan, I.~Bulu, and M.~Lon{\textbackslash}vcar.
\newblock Broadband waveguide {QED} system on a chip.
\newblock {\em Physical Review A}, 80(1):011810, 2009.

\bibitem{krishnamoorthy2012topological}
Harish N.~S Krishnamoorthy, Zubin Jacob, Evgenii Narimanov, Ilona Kretzschmar,
  and Vinod~M Menon.
\newblock Topological transitions in metamaterials.
\newblock {\em Science}, 336(6078):205--209, April 2012.

\bibitem{lezec2002beaming}
H.~J. Lezec, A.~Degiron, E.~Devaux, R.~A. Linke, L.~Martin-Moreno, F.~J.
  Garcia-Vidal, and T.~W. Ebbesen.
\newblock Beaming light from a subwavelength aperture.
\newblock {\em Science}, 297(5582):820--822, 2002.

\bibitem{de2012tailoring}
N.P. de~Leon, B.J. Shields, C.L. Yu, D.E. Englund, A.V. Akimov, M.D. Lukin, and
  H.~Park.
\newblock Tailoring light-matter interaction with a nanoscale plasmon
  resonator.
\newblock {\em Physical Review Letters}, 108(22):226803, 2012.


\bibitem{JohnsonAndChristy} PB Johnson and RW Christy. Optical Constants of the Noble Metals. {\it Physical Review B} {\bf 6}, 1972

\bibitem{NanoHub} Xingjie Ni, Zhengtong Liu, and Alexander V. Kildishev. {\it Nanohub} PhotonicsDB: Optical Constants, 2010.

\bibitem{ChirpedGratings} J.S. Bouillard, S. Vilain, W. Dickson, G.A. Wurtz, and A.V. Zayats.
\newblock{Broadband and broadangle SPP antennas basedon plasmonic crystals with linear chirp.}
\newblock{\em Nature: Scientific Report} {\bf 2}:829, 2012

\bibitem{ChirpedGratings2}
Qiaoqiang Gan and Filbert J. Bartoli.
\newblock Surface dispersion engineering of planar plasmonic chirped grating for complete visible rainbow trapping.
\newblock {\em Applied Physics Letters} {\bf 98}, 251103, 2011


\end{thebibliography}
\bibliographystyle{unsrt}


\end{document}